\documentclass[10pt]{article}
\usepackage[square,authoryear]{natbib}
\usepackage{graphicx}
\usepackage{stackrel}
\usepackage{cancel}

\usepackage{epstopdf,framed}
\usepackage{pdfsync}
\usepackage{amscd}
\usepackage{mathrsfs}

\RequirePackage{color}
\usepackage{amsmath,bm}
\usepackage{amssymb}
\usepackage{amsfonts}
\usepackage{url}

\begin{document}
\newtheorem{remark}[theorem]{Remark}

\title{Dirac structures and port-Lagrangian systems in thermodynamics}
\vspace{-0.2in}

\newcommand{\todoFGB}[1]{\vspace{5 mm}\par \noindent
\framebox{\begin{minipage}[c]{0.95 \textwidth} \color{red}FGB: \tt #1
\end{minipage}}\vspace{5 mm}\par}

\date{}
\author{\hspace{-1cm}
\begin{tabular}{cc}
Hiroaki Yoshimura & Fran\c{c}ois Gay-Balmaz
\\   School of Science and Engineering & CNRS, LMD, IPSL
\\  Waseda University & Ecole Normale Sup\'erieure 
\\  Okubo, Shinjuku, Tokyo 169-8555, Japan & 24 Rue Lhomond 75005 Paris, France 
\\ yoshimura@waseda.jp & gaybalma@lmd.ens.fr \\
\end{tabular}\\\\
}

\maketitle
\vspace{-0.3in}

\begin{center}
\abstract{In this paper, we introduce the notion of {\it port-Lagrangian systems} in nonequilibrium thermodynamics, which is constructed by 
generalizing the notion of port-Lagrangian systems for nonholonomic mechanics proposed in \cite{YoMa2006c}, where the notion of interconnections is described in terms of Dirac structures. The notion of port-Lagrangian systems in nonequilibrium thermodynamics is deduced from the variational formulation of nonequilibrium thermodynamics developed in \cite{GBYo2017a,GBYo2017b}. It is a type of Lagrange-d'Alembert principle associated to a specific class of nonlinear nonholonomic constraints, called phenomenological constraints, which are associated to the entropy production equation of the system. To these phenomenological constraints are systematically associated variational constraints, which need to be imposed on the variations considered in the principle.  In this paper, by specifically focusing on the cases of simple thermodynamic systems with constraints, we show how the interconnections in thermodynamics can be also described by Dirac structures on the Pontryagin bundle as well as on the cotangent bundle of the thermodynamic configuration space. Each of these Dirac structures is induced from the variational constraint. Furthermore, the variational structure associated to this Dirac formulation is presented in the context of the Lagrange-d'Alembert-Pontryagin principle. We illustrate our theory with some examples such as a cylinder-piston with ideal gas as well as an LCR circuit with entropy production due to a resistor. 
}
\end{center}

\tableofcontents

\section{Introduction}

\textit{Dirac structures} are geometric objects that unify and generalize the notions of symplectic and Poisson structures (\cite{CoWe1988,Cour1990}). It is well known that Dirac structures appropriately represent the so-called {\it interconnections} in physical systems such as those appearing in electric circuits and nonholonomic systems, see, \cite{VaMa1994,BC1997}. In particular, the notion of {\it port-Hamiltonian systems} was developed in the context of Dirac systems defined from Poisson brackets with applications to a large class of controlled physical systems in, for instance, \cite{VaMa1995}. Further, it was shown in \cite{EbMaVa2005, EbMaVa2007} that the notion of port-Hamiltonian system can be extended to thermodynamics in the context of contact geometry. In conjunction with port-Hamiltonian systems, a variational approach to thermodynamics was developed in \cite{MKr2013} in the context of a contact Hamiltonian with ports.

On the Lagrangian side, the theory of \textit{Lagrange-Dirac dynamical systems} in mechanics was proposed by \cite{YoMa2006a, YoMa2006b} and the associated port-Lagrangian systems, defined as {\it Lagrangian systems with external ports} through which energy flow is exchanged, were illustrated with the case of electric circuits (\cite{YoMa2006c, YoMa2008}). Regarding thermodynamics, a \textit{Lagrangian variational formulation} for the nonequilibrium thermodynamics of finite dimensional and continuum systems has been developed by \cite{GBYo2017a,GBYo2017b} following Stueckelberg's axiom of nonequilibrium thermodynamics (\cite{StSc1974}). This variational formulation extends the Hamilton principle of classical mechanics to include irreversible processes such as friction, heat, and mass transfer. In this approach, the entropy production equation is interpreted as a \textit{phenomenological constraint}, to which is systematically associated a \textit{variational constraint}, thanks to the concept of \textit{thermodynamic displacements}. The equations are obtained by a type of Lagrange-d'Alembert principle with nonlinear nonholonomic constraints. It was then clarified that the geometric structure underlying this Lagrangian formulation is a Dirac structure induced from these constraints, as shown in \cite{GBYo2018b} for simple and closed systems. The \textit{Lagrange-Dirac thermodynamic system} associated to this Dirac structure is equivalent to the system of evolution equations obtained from the variational formulation.
 
In this paper, we show how {\it port-Lagrangian systems in thermodynamics} can be constructed in the context of Dirac structures.
In particular, we show how the interconnection structure can be modeled by an induced Dirac structure in nonequilibrium thermodynamics, where we consider a class of nonlinear nonholonomic constraints due to the irreversible processes as well as nonholonomic mechanical constraints\footnote{In this paper we use the terminology \textit{mechanical constraints} for kinematic constraints in the mechanical part to distinguish from the constraints due to the irreversible processes.}.

We also present the variational structure associated to the Dirac formulation given by the \textit{Lagrange-d'Alembert-Pontryagin principle} on the Pontryagin bundle of the thermodynamic configuration space. Finally, we illustrate our theory with the case of a piston-cylinder system with ideal gas and the case of an LCR system with entropy production. We also briefly mention the case of \textit{open systems} that includes heat and matter power exchange with the exterior, where we extend the notion of ``port" to the general one through which the external power is described by the product of thermodynamic fluxes and affinities associated with the heat and matter transfers in addition to the paring of the external forces and velocities. 

\section{port-Lagrangian systems}

In this section we quickly review from \cite{YoMa2006a,YoMa2006b,YoMa2006c,YoMa2008} the theory of Lagrange-Dirac systems with external ports, also referred to as {\it port-Lagrangian systems}.

\subsection{Dirac structures and interconnection}

We first recall the definition of a Dirac structure, see \cite{CoWe1988}.

Let $V$ be a vector space, let $\langle\cdot \, , \cdot\rangle$
be the natural paring between $V$ and its dual space $V^{\ast}$, and consider the symmetric paring on $V \oplus V^{\ast}$ defined by
\begin{equation*}
\langle \! \langle (v,\alpha),
(\bar{v},\bar{\alpha}) \rangle \!  \rangle
=\langle \alpha, \bar{v} \rangle
+\langle \bar{\alpha}, v \rangle,
\end{equation*}
for $(v,\alpha), (\bar{v},\bar{\alpha}) \in V \oplus V^{\ast}$.
A \textit{linear Dirac structure} on $V$ is a subspace $D \subset V \oplus
V^{\ast}$ such that
$D=D^{\perp}$, where $D^{\perp}$ is the orthogonal
of $D$ relative to the pairing
$\langle \! \langle \,,  \rangle \!  \rangle$.

Let $M$ be a smooth manifold and let $TM \oplus T^{\ast}M$ denote the Pontryagin bundle over $M$, defined as the direct sum of the tangent and cotangent bundle of $M$. In this paper, we shall call a subbundle $ D \subset TM \oplus T^{\ast}M$ a \textit{Dirac structure} on $M$, if $D(m)$ is a linear Dirac structure on the vector space $T_{m}M$ at each point $m \in M$.  

It is well known that such a Dirac structure represents an {\it interconnection} with multi-ports (see, \cite{VaMa1995, YoMa2006a}).

For instance,  the interconnection structure due to Kirchhoff's current and voltage constraints, namely, KCL and KVL, can be modeled by a Dirac structure.

In this paper, we will show how  the interconnection in thermodynamics can be constructed by a Dirac structure as in mechanics.

\subsection{Dirac formulations for port-Lagrangian systems}

Let $Q$ be an $n$-dimensional configuration manifold and let $TQ$ and $T^\ast Q$ be the tangent and cotangent bundles over $Q$ with local expressions $(q,v)$ and $(q,p)$, respectively. We consider a Lagrangian $L:TQ\rightarrow\mathbb{R}$, possibly degenerate, as well as a mechanical constraint distribution $\Delta_Q$ on $Q$, which may be nonholonomic.

In the context of Lagrangian systems, there are two associated Dirac formulations; one given on the \textit{Pontryagin bundle} $P=TQ \oplus T^\ast Q$ and one on the \textit{cotangent bundle} $T^*Q$. We shall start with the Dirac formulation on $P$.

\paragraph{Dirac dynamical systems on the Pontryagin bundle}

The Dirac structure $D_{\Delta_P}\subset TP\oplus T^*P$ associated to $\Delta_Q$ is defined at each $\mathrm{x}=(q,v,p) \in P$ by
\begin{align*}
D_{\Delta_P}(\mathrm{x})
& =\big\{ (V, \alpha) \in T_{\mathrm{x}}P \times
T^{\ast}_{\mathrm{x}}P \mid V \in
\Delta_{P}(\mathrm{x}),   \nonumber\\
&\qquad \mbox{and} \;
\langle\alpha,W\rangle = \Omega_P(\mathrm{x}) (V,W), \;\forall \;W \in \Delta_{P}(\mathrm{x})\big\},
\end{align*}
where $\Omega_P=\pi_{(P, T^\ast Q)}^{\ast}\Omega$ is the presymplectic form on $P$ induced by the canonical form $\Omega$ on $T^*Q$ and
\[
\Delta_{P}=(T\pi_{(P,Q)})^{-1}(\Delta_Q)
\]
is the distribution on $P$ induced by the distribution $\Delta_Q$ on $Q$. 
Here $\pi_{(P,Q)}: P \to Q$ and $\pi_{(P, T^\ast Q)}: P \to T^\ast Q$ are the natural projections, locally given by $\pi_{(P,Q)}(q,v,p)=q$ and $\pi_{(P, T^\ast Q)}(q,v,p)=(q,p)$.

Let us assume that an external force field $F^{\rm ext}: TQ \to T^\ast Q$ is given through the external energy port. We define the associated horizontal one-form $\mathcal{F}^{\rm ext}$ on $P$ as follows:
\[
\langle \mathcal{F}^{\rm ext}(\mathrm{x}) , W_{\mathrm{x}}\rangle= \left\langle F^{\rm ext}(q,v), T_\mathrm{x}\pi_{(P,Q)}(W_{\mathrm{x}}) \right\rangle,
\]
for all $\mathrm{x}=(q,v,p) \in P$ and all $W_{\mathrm{x}} \in T_{\mathrm{x}}P$. Locally it reads
\begin{equation}\label{F_P}
 \mathcal{F}^{\rm ext}(q,v,p)= \left(q,v,p, F^{\rm ext}(q,v),0 ,0\right).
\end{equation}

From the given Lagrangian $L$, we define the generalized energy $E$ on the Pontryagin bundle $P$ by
\[
E(q,v,p)=\langle p, v \rangle -L(q,v).
\]

Given $E$, $D_{\Delta_P}$, and $\mathcal{F}^{\rm ext}$ constructed from $L$, $\Delta_Q$, and $F^{\rm ext}$, the associated \textit{port-Lagrangian system} for a curve $\mathrm{x}(t)=(q(t),v(t),p(t)) \in P$ is defined as:
\begin{equation}\label{PLS}
\big((\mathrm{x}(t),\dot{\mathrm{x}(t)}), \mathbf{d}E(\mathrm{x}(t))-\mathcal{F}^{\rm ext}(\mathrm{x}(t))\big) \in D_{\Delta_P}(\mathrm{x}(t)).
\end{equation}
Using local coordinates, we get the Lagrange-d'Alembert-Pontryagin equations:
\begin{equation}\label{LDAPeqn}
\begin{split}
&p =\frac{\partial L}{\partial v }(q,v), \;\; \dot q =v \in \Delta_Q(q), \;\;  \\[3mm]
&\dot p - \frac{\partial L}{\partial q}(q,v)-{F}^{\rm ext}(q,v)
\in \Delta_Q(q)^{\circ},
\end{split}
\end{equation}
where $\Delta_Q^\circ\subset T^*Q$ is the annihilator of $\Delta_Q$ in $T^*Q$.

\paragraph{Energy balance equation}

Along a solution curve $\mathrm{x}(t)=(q(t), v(t),p(t))\in P$ of system \eqref{PLS}, or equivalently \eqref{LDAPeqn}, the energy balance equation holds as:
\[
\frac{d}{dt} E (q(t), v(t), p(t))
       =\left<{F}^{\rm ext}(v(t)) , \dot{q}(t) \right>,
\]
as a direct computation shows.

\paragraph{Induced Dirac structures on the cotangent bundle}

Consider as before a kinematic constraint $\Delta_Q$ on $Q$, and define the distribution $\Delta_{T^*Q}$ on $T^{\ast}Q$ by
\[
\Delta_{T^{\ast}Q}
:=( T\pi_{Q})^{-1} \, (\Delta_{Q}) \subset TT^{\ast}Q,
\]
where $\pi_{Q}:T^{\ast}Q \to Q$ is the canonical projection and 
$T\pi_{Q}:TT^{\ast}Q \to TQ$ its tangent map.

The Dirac structure $D_{\Delta_Q}\subset TT^*Q\oplus T^*T^{\ast}Q$ induced from $\Delta_Q$ is defined for each $z=(q,p) \in
T^{\ast}Q$ by
\begin{align*}
D_{\Delta_Q}(z)
& =\{ (v, \alpha) \in T_{z}T^{\ast}Q \times
T^{\ast}_{z}T^{\ast}Q \mid v \in
\Delta_{T^{\ast}Q}(z),   \nonumber\\
&\qquad\mbox{and} \;
\langle\alpha,w\rangle= \Omega(z) (v,w),\;\forall\; w \in \Delta_{T^{\ast}Q}(z)\},
\end{align*}
where $\Omega$ is the canonical symplectic form on $T^*Q$.

Letting points in $TT^*Q$ and $T ^{\ast} T ^{\ast} Q $ be locally denoted by $(q,p,\dot q, \dot p)$ and $( q,p, \alpha, u )$, where $\alpha$ is a covector and $w $ is a vector, the local expression for the Dirac structure induced from $\Delta_Q$ is
\begin{equation}\label{local_induced_Dirac}
\begin{aligned}
D_{\Delta_Q}(q,p)  
  & =
\big\{( (q,p, \dot{q}, \dot{p}), (q,p, \alpha, u) ) \mid
\dot{q} \in \Delta _Q(q),\\
&\qquad\qquad\quad\; u = \dot{q},\;\mbox{and} \;\alpha +\dot{p} \in \Delta_Q(q)^{\circ}\big\}.
\end{aligned}
\end{equation}

\paragraph{Lagrange-Dirac systems on $T^\ast Q$} 

Consider the differential $ \mathbf{d} L:TQ \rightarrow T^{\ast}TQ$ of the Lagrangian $L$ and the canonical symplectomorphism $ \gamma_{Q}:T^{\ast}TQ \rightarrow T^{\ast}T^{\ast}Q$ given locally by $(q,\delta{q},\delta{p},p) \mapsto (q,p,-\delta{p},\delta{q})$. Following \cite{YoMa2006a}, the Dirac differential of $L$ is defined by 
\[
\mathbf{d}_{D} L:= \gamma_{Q} \circ \mathbf{d} L :TQ \rightarrow T^{\ast}T^{\ast}Q.
\]
Locally it reads
\begin{equation}\label{dDL}
\mathbf{d}_{D} L(q,v)= \left(q,\frac{\partial L}{\partial v}, - \frac{\partial L}{\partial q},  v \right).
\end{equation}

Given the external force $F^{\rm ext}: TQ \to T^\ast Q$, we define the map
\[
\widetilde{F}^{\rm ext}: TQ \to T^\ast T^\ast Q
\]
such that $\widetilde{F}^{\rm ext}(y)\in T^*_{z}T^*Q$ for $y=(q,v)$ and $z=(q,\frac{\partial L}{\partial v})$ and 
\[
\left\langle \widetilde{F}^{\rm ext}(y) , W\right\rangle = \left\langle F^{\rm ext}(y), T_z\pi_{Q}(W) \right\rangle,
\]
for all $W_z \in T_zT^*Q$.
Locally it reads
\begin{equation}\label{tildeF}
\widetilde{F}^{\rm ext}(q,v)= \left(q,\frac{\partial L}{\partial v},F^{\rm ext}(q,v),0 \right).
\end{equation}

The associated \textit{port-Lagrangian system} is given by the following {Lagrange-Dirac dynamical system} (or implicit Lagrangian system) for the curves $y(t)= (q(t),v(t))$ and $z(t)=(q(t),p(t))$:
\begin{equation}\label{LDirac_system} 
\left( (z(t),\dot{z}(t)), \mathbf{d}_{D} L(y(t))-\widetilde{F}^{\rm ext}(y(t))\right)  \in D_{ \Delta _Q }(z(t)).
\end{equation}

Using \eqref{local_induced_Dirac}, \eqref{dDL}  and \eqref{tildeF}, it follows that the port-Lagrangian system \eqref{LDirac_system} is equivalent to the Lagrange-d'Alembert-Pontryagin equations \eqref{LDAPeqn}.
 
\section{Port-Lagrangian systems in thermodynamics}

A \textit{discrete thermodynamic system} $ \boldsymbol{\Sigma} $ is a collection $ \boldsymbol{\Sigma} = \cup_{A=1}^N  \boldsymbol{\Sigma}_A $ of a finite number of interacting simple thermodynamic systems $ \boldsymbol{\Sigma} _A $. By definition, a \textit{simple thermodynamic system} is a macroscopic system for which one (scalar) thermal variable and a finite set of mechanical variables are sufficient to describe entirely the state of the system. The first law of thermodynamics, \cite{StSc1974}, asserts that for every system there exists an extensive state function, the energy, which satisfies
\begin{equation*}\label{law1_explicit}
\frac{d}{dt} E  =P^{\rm ext}_W+P^{\rm ext}_H+P^{\rm ext}_M,
\end{equation*}
where $ P^{\rm ext}_W$ is the power associated to the work done on the system, $P^{\rm ext}_H$ is the power associated to the transfer of heat into the system, and $P^{\rm ext}_M$ is the power associated to the transfer of matter into the system. In particular, a system in which $P^{\rm ext}_M \ne 0$ is called \textit{open}.


\subsection{Variational and geometric settings in thermodynamics}
\paragraph{Simple and adiabatically closed systems.}
 
Let $Q$ be the configuration manifold of the mechanical part of the thermodynamic system and let $\mathcal{Q}=Q\times \mathbb{R}$ be the thermodynamic configuration manifold for the simple system, where $\mathbb{R}$ is the space containing the entropy variable of the system $S$.  The Lagrangian of a simple thermodynamic system is given by a function
\[
L: TQ \times \mathbb{R}  \rightarrow \mathbb{R} , \quad (q, v, S) \mapsto L(q, v, S).
\]
Suppose that exterior and friction forces are given by the fiber preserving maps $F^{\rm ext}, F^{\rm fr}:TQ\times \mathbb{R} \rightarrow T^* Q$ and consider the adiabatically closed case, in which there exist no external heat power supply $P^{\rm ext}_H(t)$ nor matter exchange $P^{\rm ext}_M(t)$ with exterior. 

\paragraph{Nonholonomic constraints of thermodynamic type.}

It follows from \cite{GBYo2017a} that the evolution equations of the thermodynamic system can be obtained by a type of Lagrange-d'Alembert principle with a specific type of nonlinear nonholonomic constraints, to which are systematically associated variational constraints imposed on the variations of the curve.

For the present case, the nonholonomic constraint reads
\begin{equation}\label{PC}
\frac{\partial L}{\partial S}(q,\dot q,S)\dot S= \left\langle F^{\rm fr}(q , \dot q , S),\dot q\right\rangle,
\end{equation}
where $\frac{\partial L}{\partial S}=-T$ is minus the temperature of the system.
Such constraints arising in thermodynamic are referred to as \textit{phenomenological constraints}.
Geometrically, the constraint \eqref{PC} defines a submanifold $C_K^{\rm th}\subset T \mathcal{Q}$, given as
\[
C_K^{\rm th}=\Big\{(q,S,\dot q, \dot S)\in T\mathcal{Q}\mid \frac{\partial L}{\partial S}(q,\dot q,S)\dot S= \left\langle F^{\rm fr}(q , \dot q , S),\dot q\right\rangle\!\Big\}
\]
The \textit{variational constraint} associated to \eqref{PC} is 
\begin{equation}\label{VC}
\frac{\partial L}{\partial S}(q,\dot q,S)\delta  S= \left\langle F^{\rm fr}(q , \dot q , S),\delta q\right\rangle.
\end{equation}
Geometrically, the constraint \eqref{VC} defines a submanifold $ C_V^{\rm th} \subset T\mathcal{Q} \times_\mathcal{Q} T\mathcal{Q}$, given as
\begin{equation}\label{C_V_thermo}
\begin{aligned}
&C_V^{\rm th}= \Big \{(q,S, v, W,\delta q, \delta S) \in  T \mathcal{Q} \times _ \mathcal{Q} T \mathcal{Q}  \;\Big|\; \\
& \hspace{2cm}\frac{\partial L}{\partial S}(q,v,S)\delta S= \left\langle F^{\rm fr}(q , v , S),\delta q \right\rangle\Big \}.
\end{aligned}
\end{equation} 
Above $T\mathcal{Q} \times_\mathcal{Q} T\mathcal{Q}\rightarrow\mathcal{Q}$ is the vector bundle whose fiber at $(q,S)\in \mathcal{Q}$ is the Cartesian product $T_{(q,S)} \mathcal{Q}\times T_{(q,S)} \mathcal{Q}$. 

The relation between the phenomenological constraint $C_K^{\rm th}$ and the variational constraint $C_V^{\rm th}$ can be geometrically written as
\begin{equation}\label{CKCV}
C_K^{\rm th}=\left\{ (q,S,\dot{q},\dot{S})\mid (q,S,\dot{q},\dot{S}) \in C_V^{\rm th}(q,S,\dot{q},\dot{S}) \right\},
\end{equation}
where $C_V^{\rm th}(q,S,\dot{q},\dot{S}):= C_V^{\rm th} \cap \big(\{(q,S,\dot q, \dot S)\} \times T_{(q,S)}\mathcal{Q}\big)$.

Constraints related in this way are called constraint of thermodynamic type. It is a relation that holds for both simple and nonsimple closed systems, \cite{GBYo2017a,GBYo2017b}.

\paragraph{Nonholonomic mechanical constraints.}

In this paper, we will further assume that the above thermodynamic system is subject to nonholonomic mechanical constraints given by a distribution $\Delta_Q \subset TQ$. We shall describe $\Delta_Q$ as
\[
\Delta_Q(q)=\left\{ (q,v) \in TQ \mid \left< \omega^r(q), v \right>=0, \; r=1,...,m<n \right\},
\]
where $\omega^r=\sum_{i=1}^n \omega^r_i dq^i$ are $m$ constraint one-forms.

Using the projection $\pi_{(\mathcal{Q},Q)}:\mathcal{Q} \to Q$, $(q,S)  \mapsto q$, we define the distribution $\Delta_{\mathcal{Q}}$ naturally induced on the thermodynamic configuration space $\mathcal{Q}$ by
\begin{equation}\label{distribution_mathcalQ}
\Delta_{\mathcal{Q}}=(T\pi_{(\mathcal{Q},Q)})^{-1}(\Delta_Q) \subset T\mathcal{Q}.
\end{equation}
The geometric object corresponding to $C^{\rm th}_V$ in \eqref{C_V_thermo} is here given the mechanical variational constraint
\begin{equation}\label{CV_mech}
C^{\rm mech}_V= T\mathcal{Q}\times_\mathcal{Q}\Delta_\mathcal{Q}.
\end{equation}

For the thermodynamic system with mechanical constraint, we thus get the variational constraint $\mathcal{C}_V$
\begin{equation}\label{both}
C_V:=C_V^{\rm th} \cap C_V^{\rm mech}  \subset T\mathcal{Q}\times_\mathcal{Q}T\mathcal{Q},
\end{equation}
which is locally described by
\begin{equation}\label{CalC_V_thermo}
\begin{aligned}
C_V\!&= \!\Big \{(q,S, v, W,\delta q, \delta S)  \;\Big|\; (q,\delta{q}) \in \Delta_Q(q) \\
& \hspace{1cm}\textrm{and} \;\;\frac{\partial L}{\partial S}(q,v,S)\delta S= \left\langle F^{\rm fr}(q , v , S),\delta q \right\rangle\Big \}.
\end{aligned}
\end{equation} 
Then, as in \eqref{CKCV}, the associated kinematic constraint $\mathcal{C}_K \subset T \mathcal{Q}$  is given by
\begin{equation*}\label{CalC_K_thermo}
\begin{aligned}
C_K\!&= \!\Big \{(q,S, \dot{q}, \dot{S})  \;\Big|\; (q,S, \dot{q}, \dot{S})\in \mathcal{C}_V(q,S, \dot{q}, \dot{S}) \Big \},
\end{aligned}
\end{equation*} 
which leads to 
\begin{equation}\label{locCalC_K_thermo}
\begin{aligned}
C_K\!&= \!\Big \{(q,S, \dot q, \dot S)  \;\Big|\; (q,\dot{q}) \in \Delta_Q(q) \\
& \hspace{1cm}\textrm{and} \;\;\frac{\partial L}{\partial S}(q,\dot{q},S)\dot S= \left\langle F^{\rm fr}(q ,\dot{q} , S),\dot{q} \right\rangle\Big \}.
\end{aligned}
\end{equation} 
The annihilator of $C_V(q,S, v, W) \subset T_{(q,S)}\mathcal{Q}$ in $T^*\mathcal{Q}$ is given by
$$
C_V(q,S, v, W)^\circ= \!\Big \{(q,S, \alpha, \mathcal{T})  \;\Big|\; 
\frac{\partial L}{\partial S}\alpha   + \mathcal{T} F^{\rm fr} \in \Delta_Q(q)^\circ
\Big \}.
$$

\subsection{Port-Lagrangian systems on the Pontryagin bundle}\label{3_2}
\paragraph{Dirac structures on the Pontryagin bundle.}
Let $\mathcal{P} =T \mathcal{Q}\oplus T ^* \mathcal{Q} $ be the Pontryagin bundle over the thermodynamic configuration space $\mathcal{Q}$. Denoting $\mathrm{x}=(q,S,v,W,p, \Lambda  )$ an element in $ \mathcal{P}$ and using the projection $\pi_{(\mathcal{P} ,\mathcal{Q})}: \mathcal{P} \rightarrow \mathcal{Q}$, $\mathrm{x}=( q, S, v, W, p,\Lambda) \mapsto ( q, S)$, we can define the distribution induced from $\mathcal{C}_V$ on $\mathcal{P}$ as 
\[
\Delta _ \mathcal{P} (\mathrm{x}):= (T _{\mathrm{x}} \pi_ {(\mathcal{P},\mathcal{Q} )} ) ^{-1}(\mathcal{C}_V(q,S,v,W)).
\]
Define the presymplectic form $ \Omega _{\mathcal{P} }= \pi _{(\mathcal{P} , T^* \mathcal{Q} )} ^\ast \Omega_{T^* \mathcal{Q} } $  on $ \mathcal{P} $, induced from the canonical symplectic form $\Omega_{T^*\mathcal{Q}}= dq^i \wedge dp_i + dS \wedge d \Lambda$ on $T^*\mathcal{Q}$.

Now consider the Dirac structure on $ \mathcal{P} $ induced from $\Delta _ \mathcal{P}$ and $ \Omega _{\mathcal{P} }$ as
\begin{equation*}\label{D_thermo_P}
\begin{aligned}
D_{ \Delta _ \mathcal{P}}(\mathrm{x})
&=\big\{ (v,  \zeta) \in T_{\mathrm{x}} \mathcal{P}  \times T^{\ast}_{\mathrm{x}}\mathcal{P}  \; \mid \;  v \in \Delta _ \mathcal{P}(\mathrm{x}), \\ 
&\qquad \mbox{and} \; \left\langle  \zeta,w \right\rangle =\Omega_{\mathcal{P} }(\mathrm{x})(v,w) , \;\forall \; w \in\Delta _ \mathcal{P}(\mathrm{x}) \big\}.
\end{aligned}
\end{equation*}

Writing locally $(\mathrm{x}, \dot{\mathrm x}) \in T\mathcal{P} $ and $ (\mathrm{x},  \zeta ) \in T^* \mathcal{P} $, where $\dot{\mathrm x}= (\dot q,\dot S,\dot v,\dot W,\dot p, \dot \Lambda  )$, and $ \zeta =(\alpha ,  \mathcal{T}  ,  \beta , \Upsilon   , u, \Psi )$, we have $\big((\mathrm{x}, \dot{\mathrm x} ),(\mathrm{x} , \zeta  )\big) \in D_{ \Delta _ \mathcal{P}}(\mathrm{x})$ if and only if
\begin{equation}\label{local_thermo}
\left\{
\begin{array}{l} 
\displaystyle\vspace{0.2cm}(\dot p + \alpha ) \frac{\partial L}{\partial S} (q,v,S)\\
\displaystyle\vspace{0.2cm}\hspace{2cm}+ (  \dot \Lambda + \mathcal{T} ) F^{\rm fr}(q,v,S) \in \Delta _Q (q)^{\circ} ,\\
\displaystyle\vspace{0.2cm}\frac{\partial L}{\partial S}(q,v,S) \dot S= \left\langle F^{\rm fr}(q,v,S), \dot q \right\rangle,\\
\displaystyle\beta =0 , \quad \Upsilon   =0, \quad u=\dot q  \in  \Delta _Q (q), \quad \Psi  =\dot S.\\
\end{array}\right.
\end{equation} 

\paragraph{Port-Lagrangian systems on $\mathcal{P} =T \mathcal{Q}\oplus T ^* \mathcal{Q} $.} Let us assume that an external force field $F^{\rm ext}: TQ \times\mathbb{R}\to T^\ast Q$ is given through the external energy port. As before, we define the associated horizontal one-form $\mathcal{F}^{\rm ext}$ on $\mathcal{P}$ as follows:
\[
\langle \mathcal{F}^{\rm ext}(\mathrm{x}) , W\rangle= \left\langle F^{\rm ext}(q,v,S), T_\mathrm{x}\pi_{(\mathcal{P},Q)}(W) \right\rangle,
\]
for all $\mathrm{x}=(q,S,v,W,p, \Lambda  ) \in  \mathcal{P}$ and all $W_{\mathrm{x}} \in T_{\mathrm{x}}\mathcal{P}$. Locally it reads
\begin{equation}\label{F_mathcalP}
 \mathcal{F}^{\rm ext}(\mathrm{x}  )= \left(q,S,v,W,p, \Lambda,  F^{\rm ext}(q,v,S),0,0,0,0,0\right).
\end{equation}

The generalized energy $ \mathcal{E} : \mathcal{P} \rightarrow \mathbb{R}  $  on the Pontryagin bundle is given here by
\begin{equation}\label{GE_thermo} 
\mathcal{E} (q,S,v,W,p, \Lambda  )= \left\langle p, v \right\rangle + \Lambda  W-L(q,v,S).
\end{equation} 

Given $\mathcal{E}$, $D_{\Delta _ \mathcal{P}}$, and $\mathcal{F}^{\rm ext}$ constructed from $L$, $C_V= C_V^{\rm th}\cap C_V^{\rm mech}$, and $F^{\rm ext}$, the associated \textit{port-Lagrangian thermodynamic system} for a curve $\mathrm{x}(t)\in \mathcal{P} $ is defined as
\begin{equation}\label{PLTS}
\big((\mathrm{x}(t), \dot{\mathrm x}(t)), \mathbf{d} \mathcal{E} (\mathrm{x}(t))-\mathcal{F}^{\rm ext}(\mathrm{x}(t)) \big) \in D_{\Delta _ \mathcal{P}}(\mathrm{x}(t)).
\end{equation}
Using \eqref{local_thermo} the system \eqref{PLTS} gives the following equations for a curve $\mathrm{x}(t)=(q(t),S(t),v(t),W(t),p(t), \Lambda (t) ) $ in local coordinates
\begin{equation*}\label{implicit_thermo_phys}
\left\{
\begin{array}{l} 
\displaystyle\vspace{0.2cm} \left( \dot p- \frac{\partial L}{\partial q}(q,v,S)-F^{\rm ext}(q,v,S) \right)  \frac{\partial L}{\partial S} (q,v,S)\\
\displaystyle\vspace{0.2cm} \quad+\left( \dot \Lambda  -\frac{\partial L}{\partial S}(q,v,S) \right) F^{\rm fr}(q,v,S) \in \Delta _Q (q)^\circ,\\
\displaystyle\vspace{0.2cm}\frac{\partial L}{\partial S}(q,v,S) \dot S= \left\langle F^{\rm fr}(q,v,S), \dot q \right\rangle,\\
\displaystyle p=\frac{\partial L}{\partial v} , \quad \Lambda    =0, \quad v=\dot q \in \Delta_Q, \quad W =\dot S.
\end{array}\right.
\end{equation*} 

Finally, using Lagrange multipliers $\mu_r,\;r=1,...,m<n$, we can rearrange the above evolution equations as
\begin{eqnarray}\label{LD_on_N} 
\left\{
\begin{array}{l} 
\displaystyle\vspace{0.2cm}  \dot p - \frac{\partial L}{\partial q}(q,v,S)-  F ^{\rm fr}(q,v,S) -F^{\rm ext}(q,v,S)=\mu_r \omega^r(q),\\
\displaystyle\vspace{0.2cm} \frac{\partial L}{\partial S} (q,v,S)\dot S= \left\langle F^{\rm fr}(q,v,S), \dot q \right\rangle,\\
\displaystyle  p=\frac{\partial L}{\partial v}(q,v,S), \quad \left< \omega^r(q),\dot q\right>=0, \;r=1,...,m<n  .\\
\end{array}\right.
\end{eqnarray} 
This is the system of equation describing the evolution of the simple and closed thermodynamic system with mechanical constraints

\paragraph{Energy balance equation.}

The energy balance along the solution curve $(q(t),S(t), v(t), p(t))$ of the port-Lagrangian thermodynamic system \eqref{LD_on_N} is computed as
\[
\frac{d}{dt}E(q,S,v,p)=\left<F^{\rm ext}(q,S,v,p),\dot{q}\right>=P_W^{\rm  ext},
\]
where $E(q,S,v,p)=\left<p,v\right>-L(q,v,S)$. This is the statement of the first law of thermodynamics for the simple closed system. Note that this relation can be equivalently written as $\frac{d}{dt}\mathcal{E}({\mathrm x})=\left<\mathcal{F}^{\rm ext}(\mathrm{x}),\dot{{\mathrm x}}\right>$.

\subsection{Port-Lagrangian systems on the cotangent bundle}

In \S\ref{3_2}, We have illustrated the port-Lagrangian systems for nonequilibrium thermodynamics of simple systems by utilizing a Dirac structure on the Pontryagin bundle $\mathcal{P}=T \mathcal{Q} \oplus T^* \mathcal{Q} $ over the thermodynamic configuration space $\mathcal{Q} $. In a similar way with the case of mechanics, there is another type of Dirac formulation of port-Lagrangian thermodynamic systems, which is based on a Lagrange-Dirac system and use a Dirac structure on the cotangent bundle $T^*\mathcal{Q}$ of the thermodynamic phase space.

\paragraph{Nonholonomic constraints of thermodynamic type.} 

Let us assume that the Lagrangian $L=L(q,v,S): TQ  \times \mathbb{R}  \rightarrow \mathbb{R}  $ is \textit{hyperregular with respect to the mechanical variables $(q,v)$}, that is, the Legendre transform
\begin{equation}\label{PLT} 
\mathbb{F}L_S: TQ   \rightarrow T^* Q, \quad (q,v) \mapsto \Big( q, \frac{\partial L}{\partial v}(q,v,S) \Big)  
\end{equation} 
is a diffeomorphism  for each fixed $S \in \mathbb{R}  $. From the variational constraint $C_V^{\rm th}\subset T\mathcal{Q} \times _\mathcal{Q}  T \mathcal{Q}$ in \eqref{C_V_thermo}, define the variational constraint $\mathcal{C}_V^{\rm th} \subset T^*\mathcal{Q} \times _\mathcal{Q}  T \mathcal{Q} $ as
\begin{equation}\label{def_CCCV} 
\mathcal{C} _V^{\rm th}(q,S,p, \Lambda ):= C_V^{\rm th}(q,S,v,W),
\end{equation} 
where $v$ on the right hand side is uniquely determined such that $ \frac{\partial L}{\partial v}(q,v,S)=p$. Note that $C_V^{\rm th}$ does not depend on $W$ in thermodynamics, thus $\mathcal{C} _V^{\rm th}$ is well-defined by \eqref{def_CCCV}. We have
\begin{equation}\label{C_V_thermo_Cot}
\begin{aligned}
&\mathcal{C}_V^{\rm th}(q,S,p, \Lambda )= \!\Big \{(q,S, \delta q, \delta S) \in  T \mathcal{Q}  \;\Big|\; \\
&\qquad\qquad -T(q,p,S) \delta S= \left\langle \mathcal{F} ^{\rm fr}(q,p,S),\delta q \right\rangle\Big \},
\end{aligned}
\end{equation} 
where $T$ and $\mathcal{F}^{\rm fr}$ are defined on $T ^\ast Q \times \mathbb{R}  $ by $T(q,p,S):= -\frac{\partial L}{\partial S}(q,v,S)$ and $\mathcal{F} ^{\rm fr}(q,p,S):= F^{\rm fr}(q,v,S)$, in which $v$ is uniquely determined from the condition $ \frac{\partial L}{\partial v}(q,v,S)=p$.

Recall that from the given nonholonomic mechanical constraint $\Delta_Q$, we consider the lifted distribution $\Delta_{\mathcal{Q}}$ on $\mathcal{Q}$ as in \eqref{distribution_mathcalQ}. On the Hamiltonian side, the analogue of \eqref{CV_mech} is $\mathcal{C}_V^{\rm mech}= T^*\mathcal{Q}\times_\mathcal{Q}\Delta_\mathcal{Q}$. 

As in \eqref{both} the variational constraint associated to the mechanical and thermodynamical constraints, is defined as
\begin{equation}\label{both_cotangent}
\mathcal{C}_V:=\mathcal{C}_V^{\rm th} \cap \mathcal{C}_V^{\rm mech}  \subset T^*\mathcal{Q}\times_\mathcal{Q}T\mathcal{Q}.
\end{equation}
Locally it is described by
\begin{equation}\label{scrC_V_thermo}
\begin{aligned}
\mathcal{C} _V\!&= \!\Big \{(q,S, p, \Lambda,\delta q, \delta S)  \;\Big|\; (q,\delta{q}) \in \Delta_Q(q) \\
& \hspace{1cm}\textrm{and} \;-T(q,p,S)\delta S= \left\langle  \mathcal{F}^{\rm fr} (q , p , S),\delta q \right\rangle\Big \}.
\end{aligned}
\end{equation} 
As in \eqref{CKCV}, the kinematic constraint $\mathcal{C} _K \subset T \mathcal{Q}$  is given from $\mathcal{C}_V$ as
\begin{equation*}\label{scrC_K_thermo}
\begin{aligned}
\mathcal{C} _K\!&= \!\Big \{(q,S, \dot{q}, \dot{S})  \;\Big|\; (q,S, \dot{q}, \dot{S})\in \mathcal{C} _V(q,S, p, \Lambda) \Big \},
\end{aligned}
\end{equation*} 
which leads to the local expression
\begin{align*}
&\mathcal{C}_K\!= \!\Big \{(q,S, \dot q, \dot S)  \;\Big|\; (q,\dot{q}) \in \Delta_Q(q) \\
& \hspace{1cm}\textrm{and} \;-T(q,p,S)\dot S= \left\langle \mathcal{F}^{\rm fr} (q , p , S),\dot{q} \right\rangle\Big \}.
\end{align*} 
The annihilator of $\mathcal{C} _V(q,S, v, W) \subset T\mathcal{Q}$ in $T^*\mathcal{Q}$ is
\begin{align*}
\mathcal{C}_V(q,S, p, \Lambda)^\circ&= \!\Big \{(q,S, \alpha, \mathcal{T})  \;\Big|\;-T\alpha   + \mathcal{T} \mathcal{F}^{\rm fr} \in \Delta_Q(q)^\circ
\Big \}.
\end{align*} 
From the variational constraint $\mathcal{C}_V$ in \eqref{scrC_V_thermo}, we define the distribution $\Delta_{T^*\mathcal{Q}}$ on $T ^* \mathcal{Q}$ as
\[
\Delta _{T^* \mathcal{Q} }(z):= (T _z\pi_ {(T^{\ast}\mathcal{Q},\mathcal{Q})} ) ^{-1}( \mathcal{C} _{V}(z)). 
\]
where $z=(q,S,p, \Lambda  ) \in  T ^* \mathcal{Q} $ and $ \pi _ {(T^{\ast}\mathcal{Q},\mathcal{Q})} : T ^* \mathcal{Q} \rightarrow \mathcal{Q} $, $ \pi _ {(T^{\ast}\mathcal{Q},\mathcal{Q})}(q,S,p, \Lambda  )=(q,S)$.

\paragraph{Induced Dirac structures on $T^* \mathcal{Q}$.} The Dirac structure $D_{\Delta _{T^* \mathcal{Q} }}\subset TT^*\mathcal{Q}\oplus T^*T^{\ast}\mathcal{Q}$ associated to $\Delta _{T^* \mathcal{Q} }$ is defined for each $z=(q,S,p,\Lambda) \in
T^{\ast}Q$ by
\begin{equation*}
\begin{aligned}
&D_{\Delta _{T^* \mathcal{Q} }}(z) 
\!=\!\big\{ (V, \beta) \in T_{z}T^*\mathcal{Q} \times T^{\ast}_{z}T^*\mathcal{Q} \, \mid \,  V \in \Delta _{T^* \mathcal{Q} }(z)  \\ 
&\qquad\qquad\text{and} \left\langle \beta,W \right\rangle =\Omega_{T^{\ast}\mathcal{Q}}(z)(V,W) , \;\forall \; W\in \Delta _{T^* \mathcal{Q} }(z) \big\}.
\end{aligned}
\end{equation*}
Writing locally $(z, \dot z) \in TT^*\mathcal{Q}$ and $(z, \zeta  )\in T^*T^*\mathcal{Q}$, where $\dot z= (\dot q,\dot S,\dot p, \dot \Lambda  )$ and $\zeta =(\alpha ,  \mathcal{T}  , u, \Psi)$, we have $\big((z, \dot z ),(z, \zeta  )\big) \in D_{\Delta _{T^* \mathcal{Q} }}(z)$ if and only if
\begin{eqnarray}\label{local_induced_Dirac_mathcal}
\left\{
\begin{array}{l} 
\displaystyle\vspace{0.2cm}-(\dot p + \alpha ) T(q,p,S)+  ( \dot \Lambda+ \mathcal{T}    )\mathcal{F} ^{\rm fr}(q,p,S) \in \Delta_Q^\circ(q),\\
\displaystyle\vspace{0.2cm}T(q,p,S) \dot S=- \left\langle \mathcal{F} ^{\rm fr}(q,p,S), \dot q \right\rangle,\\
\displaystyle  \dot q=u  \in \Delta_Q, \quad \Psi  =\dot S.
\end{array}\right.
\end{eqnarray}

\paragraph{Port-Lagrangian system on $T^* \mathcal{Q}$.} Given the Lagrangian $L:TQ\times \mathbb{R}\rightarrow\mathbb{R}$, we lift it onto $T\mathcal{Q}$, denoted $\widetilde{L}: T \mathcal{Q}\rightarrow\mathbb{R}$, so that its Dirac differential can be defined similarly as before, namely
\[
\mathbf{d} _D\widetilde{L}:= \gamma _ \mathcal{Q} \circ \mathbf{d} \widetilde{L}: T \mathcal{Q}\rightarrow T^*T^*\mathcal{Q}
\]
which reads locally
\begin{equation}\label{dDL_mathcal}
\mathbf{d} _D\widetilde{L}(q,S,v,W)= \Big( q, S,\frac{\partial L}{\partial v},0, - \frac{\partial L}{\partial q},- \frac{\partial L}{\partial S}, v, W \Big).
\end{equation}

Given the external force $F^{\rm ext}: TQ\times\mathbb{R} \to T^\ast Q$, we define the map
\[
\widetilde{F}^{\rm ext}: T\mathcal{Q} \to T^\ast T^\ast \mathcal{Q}
\]
such that $\widetilde{F}^{\rm ext}(y)\in T^*_{z}T^*\mathcal{Q}$ for $y=(q,v,S,W)$ and $z=(q,S,\frac{\partial L}{\partial v},0)$ and 
\[
\left\langle \widetilde{F}^{\rm ext}(y) , W\right\rangle = \left\langle F^{\rm ext}(y), T_z\pi_{(T^\ast \mathcal{Q}, Q)}(W) \right\rangle,
\]
for all $W_z \in T_zT^*Q$.
Locally it reads
\begin{equation}\label{tildeF_mathcal}
\widetilde{F}^{\rm ext}(q,S,v,W)\!=\! \Big(q,S,\frac{\partial L}{\partial v},0,F^{\rm ext}(q,v,S),0 ,0,0\Big).
\end{equation}

The associated \textit{port-Lagrangian system} for the simple adiabatically closed system is given by the following {Lagrange-Dirac dynamical system} (or implicit Lagrangian system) for the curves $y(t)= (q(t),S(t),v(t),W(t))$ and $z(t)=(q(t),S(t),p(t),\Lambda(t))$:
\begin{equation}\label{LDirac_system_thermo} 
\left( (z(t),\dot{z}(t)), \mathbf{d}_{D} \widetilde{L}(y(t))-\widetilde{F}^{\rm ext}(y(t))\right)  \in D_{ \Delta _\mathcal{Q} }(z(t)).
\end{equation} 

Using \eqref{local_induced_Dirac_mathcal}, \eqref{dDL_mathcal}  and \eqref{tildeF_mathcal}, it follows that the port-Lagrangian system \eqref{LDirac_system_thermo} is equivalent to the system
\begin{eqnarray}\label{LD_on_CotBundle} 
\left\{
\begin{array}{l} 
\displaystyle\vspace{0.2cm} -\left(  \dot p- \frac{\partial L}{\partial q}(q,v,S) \right)  T(q,p,S)\\
\hspace{.5cm}+ \left( \dot \Lambda -\frac{\partial L}{\partial S}(q,v,S) \right)  \mathcal{F} ^{\rm fr}(q,p,S)\in \Delta_Q^\circ(q),\\[5mm]
\displaystyle\vspace{0.2cm}T(q,S,p) \dot S= -  \left\langle \mathcal{F} ^{\rm fr}(q,p,S), \dot q \right\rangle,\\
\displaystyle  \dot q=v \in \Delta_Q(q), \;\; W  =\dot S, \;\; p=\frac{\partial L}{\partial v}(q,v,S), \;\; \Lambda =0,\\
\end{array}\right.
\end{eqnarray} 
where the last two equalities come from the fact that $(q,S,p, \Lambda , \dot q, \dot S,\dot p, \dot\Lambda )$ and $\mathbf{d} _D\widetilde{L}(q,S,v,W)$ must both belong to the fibers at the same point $(q,S,p, \Lambda ) \in T^{\ast}\mathcal{Q}$.
Recalling that $T(q,p,S)= -\frac{\partial L}{\partial S}(q,v,S)$ and $\mathcal{F} ^{\rm fr}(q,p,S)= F^{\rm fr}(q,v,S)$, it follows that \eqref{LD_on_CotBundle} is equivalent to \eqref{LD_on_N}, and yield the evolution equation for the simple closed thermodynamic system with mechanical constraints.

\section{Examples}

\subsection{A piston-cylinder system}
\begin{figure}[h]
\begin{center}
\includegraphics[scale=.9]{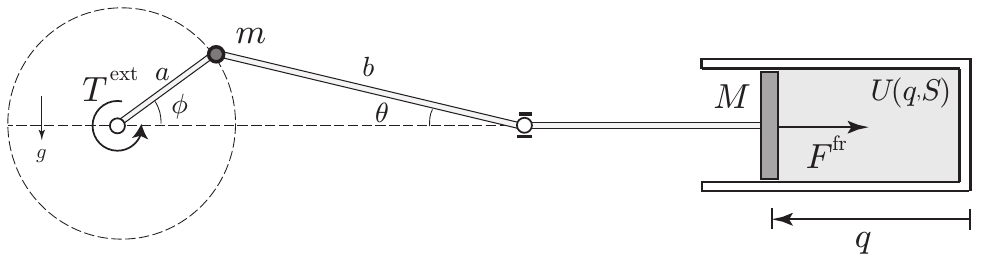} 
\caption{A piston-cylinder system with gas and friction} 
\label{fig:piston_cylinder}
\end{center}
\end{figure}

Consider the piston-cylinder system in Fig.\ref{fig:piston_cylinder}, which was considered by \cite{So1964}.
The piston has mass $M$ which is connected to the shaft with mass $m$ by massless links with lengths $a$ and $b$. Inside the cylinder, an ideal gas is contained with internal energy $U(S,V,N)$. The internal energy of the system is $U_{\rm gas}(q,S):=U(S,V=Aq,N_0)$ where $N_0$ is the constant number of moles, $V=Ax$ is the volume, and $A$ is the area of the cylinder. We assume that there is a mechanical power exchange due to an external torque $T^{\rm ext}$ acting along the shaft, but we assume no heat power or mass exchanges with exterior. We also assume that the gravitational acceleration $g$  is acting vertically  downward.

Let $\mathcal{Q}=Q \times \mathbb{R}$ be the thermodynamic configuration space, where $Q= \mathbb{R} \times S^1 \ni (q,\phi)$ is the configuration space for the mechanical part. 
The Lagrangian $L: TQ \times \mathbb{R} \to \mathbb{R} $ is given by 
\[
L(q,\phi, v_q, v_{\phi}, S)=\frac{1}{2}M v_q^2 +\frac{1}{2}m a^2 v_{\phi}^2 -U(q,\phi, S),
\]
where $U(q,\phi, S)=U_{\rm gas}(q,S)+mga \sin\phi$.

The mechanical constraint, see \cite{So1964}, is given as
\[
\Delta_Q=\{(q, \phi,  \delta{q}, \delta{\phi}) \mid  \delta{q}+\alpha(\phi) \delta \phi=0  \},
\]
where 
\[
\alpha(\phi)=a \sin \phi  \left( 1+\frac{\frac{a}{b} \cos \phi}{\sqrt {1-\left(\frac{a}{b}\right)^2\sin ^2 \phi}}\right). 
\]
It follows from \eqref{CalC_V_thermo} that the variational constraint $C_V \subset T\mathcal{Q} \times_\mathcal{Q} T\mathcal{Q}$ is given by
\[\begin{aligned}
C_V\!&= \!\Big \{(q,\phi, S,v_q,v_\phi, v_S, \delta{q}, \delta{\phi}, \delta{S})  \;\Big|\; \delta{q}+\alpha(\phi) \delta \phi=0 \\
& \;\;\textrm{and} \;\frac{\partial L}{\partial S}(q,\phi, v_q, v_{\phi}, S)\delta S= \left\langle F^{\rm fr}(q,\phi, v_q, v_{\phi}, S),\delta q \right\rangle\Big \}.
\end{aligned}
\]
From \eqref{LD_on_N}, we get the coupled mechanical and thermal evolution equations of the piston-cylinder system as follows
\begin{eqnarray*}
\left\{
\begin{array}{l} 
\displaystyle\vspace{0.2cm}  \dot p_q - p(x,S)A-  r \dot{q} =\mu,\quad \dot {q}=v_q,\\
\displaystyle\vspace{0.2cm}  \dot p_{\phi} +mga \sin \phi =\alpha(\phi) \mu + T^{\rm ext}, \quad \dot{\phi}=v_{\phi}, \\
\displaystyle\vspace{0.2cm} \frac{\partial L}{\partial S} (q,v,S)\dot S= \left\langle r \dot q, \dot q \right\rangle,\\
\displaystyle  p_q=Mv_q, \quad p_{\phi}=ma^2\dot{\phi} \quad \dot{q}=-\alpha(\phi) \dot \phi
\end{array}\right.
\end{eqnarray*} 
In the above, $F^{\rm fr}(q,v,S)=-r v$, where $r>0$ denotes the friction coefficient factor and $\frac{\partial U_{\rm gas}}{\partial q}= -{\sf p}(x,S)A$, where ${\sf p}(x,S)$ is the pressure of the ideal gas.

\subsection{Electric circuits with resistors}

Next we consider an L-C-R electric circuit with voltage source $V$, where we consider the entropy production due to the resistor $R$. The thermodynamic configuration space of this system is given by $\mathcal{Q}=Q \times \mathbb{R}$ with local coordinates $(q,S)$, where  $Q=\mathbb{R}^4$ denotes the configuration space for circuits with local coordinates for branch charges $q=(q^1,q^2,q^3,q^4)=(q_L,q_C, q_V, q_R) \in Q$. Note that $T_qQ$ is the space of branch currents with local coordinates $f=(f^1,f^2,f^3,f^4)=(f_L,f_C, f_V, f_R)$ and also that $T^\ast _qQ$ is the space of the branch voltages with $e=(e_1,e_2,e_3,e_4)=(e_L,e_C, e_V, e_R)$. 

\begin{figure}[h]
\begin{center}
\includegraphics[scale=.6]{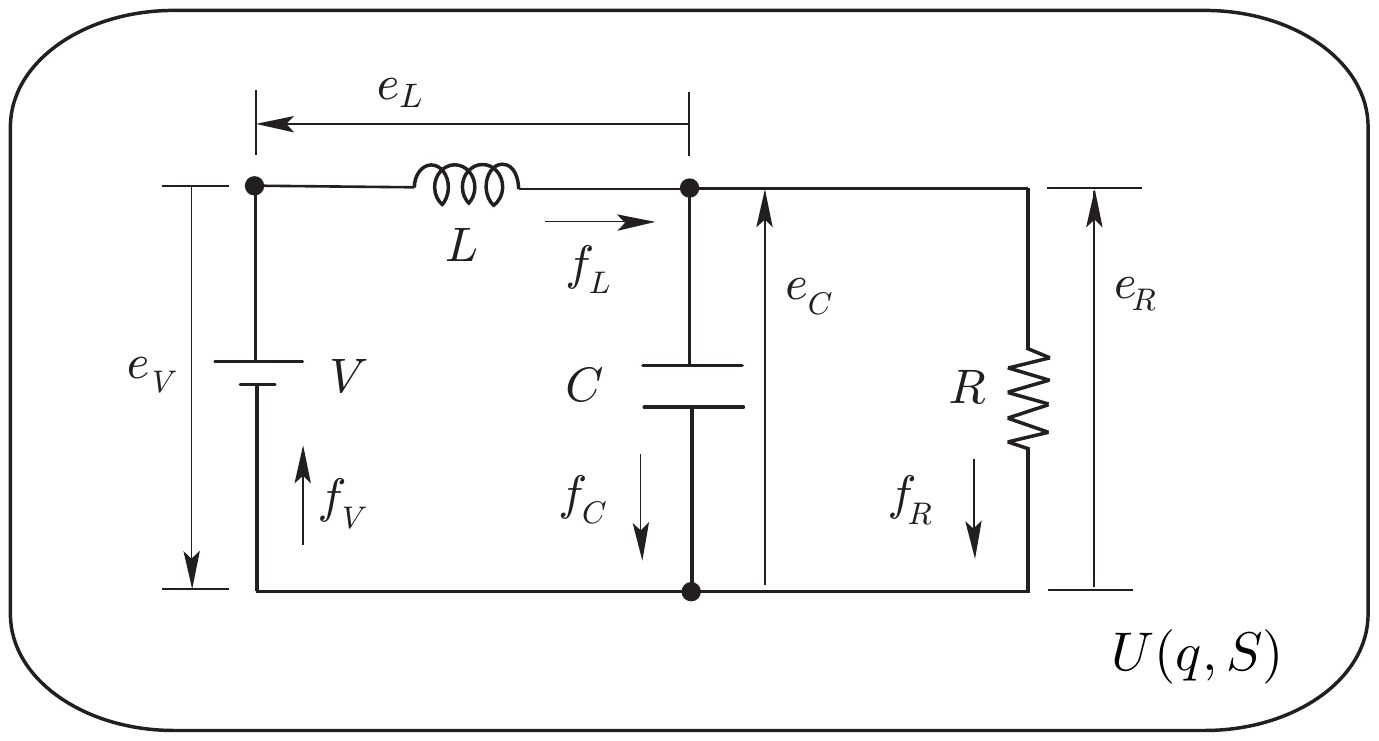}
\caption{An electric circuit with entropy production}
\label{FluidPiston2}
\end{center}
\end{figure}

The Lagrangian $L:TQ \times \mathbb{R}$ of this circuit  is given by
$$
L(q,v,S):=K_L(q,v)-U(q, S).
$$
In the above, $K_L(q,v)=\frac{1}{2}Lv_L^2$ denotes the energy due to flux linkages and $U(q, S)=U_{\rm C}(e_c)+U_{\rm int}(S)$ the potential energy consisting of the electric potential $U_{\rm C}(e_c)=\frac{1}{2C}e_c^2$ due to the capacitor and the internal energy of the system $U_{\rm int}(S)$. The KCL (Kirchhoff Current Law) constraints on the currents is given by a subspace $\Delta_Q \subset TQ$, which we shall call the {\it constraint KCL space}, defined, for each $q \in Q$, by
\[
\Delta_Q(q)=\{ f \in T_qQ \mid \langle \omega^{a}, f \rangle =0, \; a=1, 2 \},
\]
where $\omega^{a}=\omega_{k}^{a} \, dq^{k}$, for $a=1,2$ and $k=1,...,4$, 
and the coefficients $\omega_{k}^{a}$ are given in matrix by
\[
\omega_{k}^{a}=
\left(
\begin{array}{cccc}
-1 & 0 & 1 & 0 \\
0 & -1 & 1 & -1 \\
\end{array}
\right).
\]
It follows from \eqref{LD_on_CotBundle}  that the evolution equations of the port-Lagrangian thermodynamic system are given by
\[
\begin{split}
&\dot{q}_{L}=f_{L}, \quad \dot{q}_{C}=f_{C_{1}},\quad\dot{q}_{V}=f_{V},\quad  \dot{q}_{R}=f_{R},\\
&\dot{p}_{L}=-\mu_{1}+\mu_2,\quad \mu_1=e_V,\quad\mu_2=R\dot{q}_R,\\
&p_{L}=L\,f_{L},\quad f_L=f_V,\quad  f_C=f_L-f_R,\\
&\dot{S}=\frac{1}{T}q_R^2,
\end{split}
\]
which eventually yield the equations of motion:
\[
\begin{split}
&L\ddot{q}_L=V+Rf_R,\quad  R\dot{q}_R=\frac{1}{C}q_C,\\
&\; \dot{q}_C=\dot{q}_L-\dot{q}_R,\quad  \dot{S}=\frac{1}{T}R\dot{q}_R^2,
\end{split}
\]
where $T=\frac{\partial U}{\partial S}$ denotes the temperature and $e_V=-V$.

\subsection{The case of open systems}\label{case_of_open_systems}

Our approach can be applied to the more general cases of open system that allow to include matter and heat exchange with exterior. For such a simple open system, we need to consider the formulation of time-dependent nonlinear nonholonomic systems, in which the thermodynamic configuration manifold is given by
\[
Y:=\mathbb{R}\times \mathcal{Q}\ni (t,x).
\]
This manifold can be interpreted as a trivial vector bundle $Y=\mathbb{R}\times\mathcal{Q}\rightarrow \mathbb{R}$, $(t,x)\mapsto t$, over the space of time $\mathbb{R}$. Further, $\mathcal{Q}$ is the \textit{extended configuration manifold} given by $\mathcal{Q}=Q \times \mathbb{R}$, where $Q$ is the configuration manifold for mechanical part of the system and $\mathbb{R}$ is the space of the entropy variable. In this setting, we consider the vector bundle $(\mathbb{R}\times T\mathcal{Q})\times_Y T Y\rightarrow Y$ over $Y$ whose vector fiber at $y=(t,x)$ is given by $T_x\mathcal{Q} \times T_{(t,x)}Y=T_x\mathcal{Q} \times (\mathbb{R}\times  T_x\mathcal{Q})$. An element in the fiber at $(t,x)$ is denoted $(t,x,v,\delta t, \delta x)$. 

In this  setting, a \textit{variational constraint} is a subset
\[
C_V \subset (\mathbb{R}\times T\mathcal{Q})\times_Y T Y,
\]
such that $C_V(t,x,v)$, defined by
\[
C_V(t,x,v):=C_V\cap \left(\{(t,x,v)\}\times T_{(t,x)}Y\right),
\]
is a vector subspace of $T_{(t,x)}Y$, for all $(t,x,v)\in \mathbb{R}\times T\mathcal{Q}$. This constraint encodes the entropy production equation including the effect of the exterior of the system, as shown in \cite{GBYo2018a}. A \textit{kinematic constraint} is by definition a submanifold
\[
C_K\subset TY.
\]
Then, we consider the \textit{covariant Pontryagin bundle} as
\[
\pi_{(\mathcal{P},Y)}: \mathcal{P}= (\mathbb{R} \times T\mathcal{Q}) \times_{Y} T^\ast Y \rightarrow Y=\mathbb{R}\times\mathcal{Q},
\]
and we can define the distribution on $\mathcal{P}$ as
\[
\Delta_{\mathcal{P}}:=\big(T\pi_{(\mathcal{P},Y)}\big)^{-1}(C_{V}) \subset T\mathcal{P}.
\]
Let us consider the canonical symplectic form $\Omega_{T^*Y}$ on $T^*Y$ given locally as $\Omega_{T^*Y}=dx^{i} \wedge dp_{i} + dt \wedge d\mathsf{p}$. Using the projection $\pi_{(\mathcal{P}, T^{\ast}Y)}:\mathcal{P} \rightarrow T^{\ast}Y$, $(t,x,v,\mathsf{p}, p)\mapsto(t,x,\mathsf{p}, p)$ onto $T^{\ast}Y$, we get the presymplectic form on the covariant Pontryagin bundle given by
$
\Omega_{\mathcal{P}}=\pi_{(\mathcal{P}, T^{\ast}Y)}^{\ast}\Omega_{T^{\ast}Y}.
$
Then, using $\Delta_{\mathcal{P}}$ and $\Omega_{\mathcal{P}}$, we can construct the Dirac structure $D_{\Delta_{\mathcal{P}}}$ on $\mathcal{P}$ as above.

\quad Given a Lagrangian $L: \mathbb{R}\times T \mathcal{Q}\rightarrow\mathbb{R}$, the \textit{covariant generalized energy} $\mathcal{E}: \mathcal{P}\rightarrow\mathbb{R}$ on the Pontryagin bundle is defined by $\mathcal{E}= \mathsf{p} + \langle p, v\rangle -  L(t,x,v)$. Finally, given an external force $F^{\rm ext}: TQ\times\mathbb{R}\rightarrow T^*Q$, we can define as above the one form $\mathcal{F}^{\rm ext}$ on $\mathcal{P}$.

Given $\mathcal{E}$, $D_{\Delta_\mathcal{P}}$, and $\mathcal{F}^{\rm ext}$, constructed from $L$, $\mathcal{C}_V$, and $F^{\rm ext}$, the associated \textit{port-Lagrangian thermodynamic system} for a curve $\mathrm{x}(t)=(t,x(t),v(t),\mathsf{p}(t), p(t)) \in \mathcal{P}$ is defined as
\begin{equation}\label{open_Dirac}
\big(\dot{\mathrm{x}}(t), \mathbf{d}\mathcal{E}({\mathrm{x}(t)})-\mathcal{F}^{\rm ext}(\mathrm{x}(t))\big) \in D_{\Delta_{\mathcal{P}}}({\mathrm{x}(t)}).
\end{equation}
Ii gives the evolution equations for the open thermodynamic system.

\subsection{Example: a piston with heat and matter exchanges}
We consider the open system, given by a piston with heat and matter exchanges as in \ref{FluidPiston2}, in which  $S$ is the entropy of the system and $N$ is the number of moles of the chemical species and the Lagrangian $L$ is a function defined on the state space $TQ\times\mathbb{R}  \times \mathbb{R}$.  We assume that the system has $A$ ports, through which species can flow out or into the system and $B$ heat sources. We denote by $\mu^a$ and $T^a$ the chemical potential and temperature at the $a$-th port and by $T^b$ the temperature of the $b$-th heat source. 

\begin{figure}[h]
\begin{center}
\includegraphics[scale=.70]{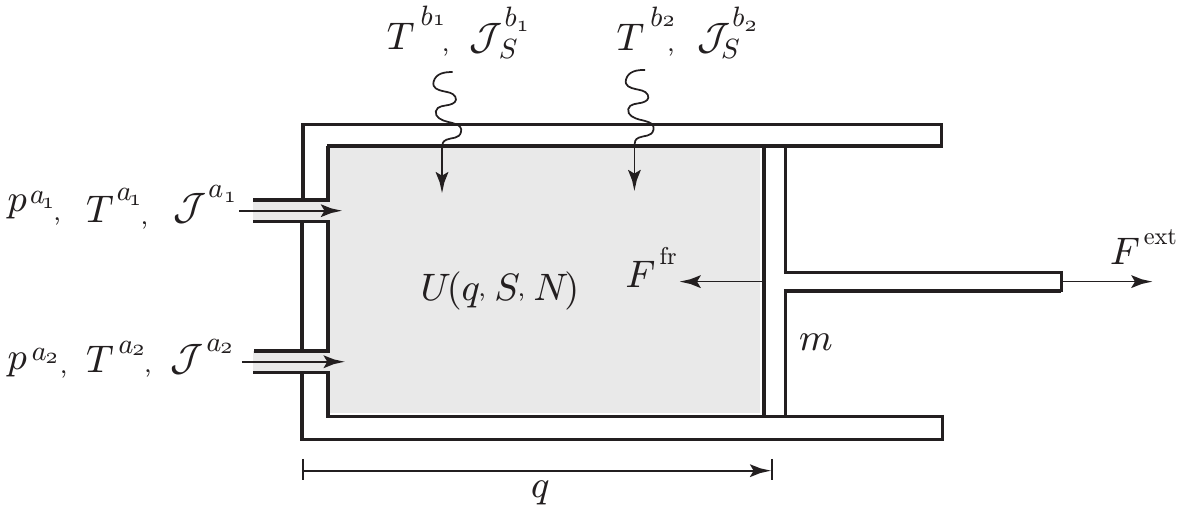}
\caption{A piston with two exterior ports and heat sources}
\label{FluidPiston2}
\end{center}
\end{figure}

The evolution equations of this system can be obtained via the port-Lagrangian system approach developed in \S\ref{case_of_open_systems}. In particular, \eqref{open_Dirac} yields the following system of evolution equations for the curves $q(t)$, $S(t)$, $N(t)$:
{\small
\begin{equation}\label{open_system}
\left\{
\begin{array}{l}
\displaystyle\frac{d}{dt}\frac{\partial L}{\partial \dot q}- \frac{\partial L}{\partial q}= F^{\rm fr} +F^{\rm ext},\quad\quad\quad  \frac{d}{dt} N= \sum_{a=1}^A \mathcal{J}^a,\\
\displaystyle\frac{\partial L}{\partial S}\Big(\dot S -\sum_{a=1}^A \mathcal{J}_S^a - \sum_{b=1}^B \mathcal{J}_S^b\Big)\\
\displaystyle\quad=  \left<F^{\rm fr},\dot q\right> -\sum_{a=1}^A\left[\mathcal{J}^a\Big(\frac{\partial L}{\partial N}+\mu^a\Big)+\mathcal{J}^a_S\Big(\frac{\partial L}{\partial S}+ T^a\Big)\right]\\[5mm]
\quad\quad-\sum_{b=1}^B \mathcal{J}^b_S\Big(\frac{\partial L}{\partial S}+ T^b\Big).
\end{array} \right.
\end{equation}
}
\!The energy balance for this system is computed as
\[
\frac{d}{dt}E=\underbrace{\left<F^{\rm ext},\dot q\right>\textcolor{white}{\sum_{b=1}^B\hspace{-0.5cm}}}_{=P^{\rm ext}_W} \;+\; \underbrace{\sum_{b=1}^B \mathcal{J}^b_S T^b}_{=P^{\rm ext}_H} \; + \; \underbrace{\sum_{a=1}^A (\mathcal{J}^a\mu^a + \mathcal{J}_S^a T^a)}_{=P^{\rm ext}_M},
\]
where the external powers are transmitted through the ports, which are given by the product of thermodynamic fluxes and affinities associated with the heat and matter transfers in addition to the paring of the external forces and velocities. 
From the last equation in \eqref{open_system}, the rate of entropy equation of the system is found as
\begin{equation}\label{entropy_open}
\dot S=I + \sum_{a=1}^A \mathcal{J}^a_S + \sum_{b=1}^B \mathcal{J}^b_S,
\end{equation}
where $I$ is the rate of internal entropy production given by
{\small
\[
\begin{aligned}
I=& \underbrace{-\frac{1}{T}\left<F^{\rm fr},\dot q\right>\textcolor{white}{\sum_{b=1}^B\hspace{-0.5cm}}}_{\text{mechanical friction}} +\underbrace{\frac{1}{T} \sum_{a=1}^A\left[\mathcal{J}^a\Big(\mu ^ a- \mu\Big)+\mathcal{J}^a_S\Big( T^a - T\Big)\right]}_{\text{mixing of matter flowing into the system}}\\
&\qquad\qquad\qquad \; + \; \underbrace{\frac{1}{T}\sum_{b=1}^B \mathcal{J}^b_S\Big(T^b- T\Big)}_{\text{heating}}.
\end{aligned} 
\]
}
\!\!We refer to \cite{GBYo2019} for the mathematical details of the formulation.

\section{Conclusions}
In this paper, we have presented the concept of port-Lagrangian systems in thermodynamic, which can be constructed with the help of Dirac structures induced from a class of \textit{nonholonomic constraints of thermodynamic type} and linear nonholonomic \textit{mechanical constraints}. We have illustrated how the interconnection structure can be modeled by such an induced Dirac structure in nonequilibrium thermodynamics. Then, we have developed the Dirac formulation by using Dirac structures on the Pontryagin bundle as well as on the cotangent bundle of the thermodynamic phase space. Finally, we have applied our theory to the examples of a piston-cylinder system with ideal gas and an LCR system with entropy production. We have also briefly presented the case of open systems that include heat and matter exchange with exterior, together with an example.

\paragraph{Acknowledgements.}
H.Y. is partially supported by JSPS Grant-in-Aid for Scientific Research (17H01097), the MEXT “Top Global University Project” and Waseda University (SR 2019C-176) and the Organization for University Research Initiatives (Evolution and application of energy conversion theory in collaboration with modern mathematics); F.G.B. is partially supported by the ANR project GEOMFLUID, ANR-14-CE23-0002-01.

\end{document}